\documentstyle[11pt,newpasp,twoside,epsf]{article}
\reversemarginpar

\begin{document}
\title{Miras and Other Cool Variables with GAIA}
\author{Michael Feast}
\affil{Astronomy Department, University of Cape Town, Rondebosch, 7701,
South Africa\\
mwf@artemisia.ast.uct.ac.za}

\begin{abstract}
A general review is given of Mira and some other cool variables stars,
concentrating on those aspects on which GAIA, and more particularly
GAIA spectroscopy and radial velocities, will have a major impact.
\end{abstract}

\section{Introduction}
The observation of Miras and other late type variables with GAIA will bring
rather special rewards, but it will also bring special problems in
analysing and interpreting the data. In reviewing some of these promises
and problems I hope it will become clear what a vital part in the success
of the mission will be played by the spectroscopy; radial velocities
and spectral data. I shall also mention ground based observations which will
be need to make full use of GAIA data. 

Miras are large amplitude, AGB variables with rather regular periods
which range from about 100 days to 1000 days or more. At the longer
periods they are often
OH/IR sources. They can be broadly divided into oxygen-rich and carbon-rich
objects (O- and C- Miras).

It has been estimated (Eyers \& Cuypers 2000) that GAIA will discover and 
measure about
150,000 Mira variables. For a significant fraction of these, GAIA will
obtain parallaxes, proper motions, radial velocities, spectral data,
narrow and broad band photometry and periods. For many of the stars 
observations will cover the whole light cycle. 

Miras are of central importance to a number of key issues in astronomy.
Amongst these are the following. (1) They appear to define the
end point of the AGB when mass-loss plays a major role in stellar evolution.
This is a time in the life of a star which is not properly
understood. (2) In old or intermediate age populations, the Miras are the
brightest individual stars and hold out the possibility of calibrating
the populations of nearby galaxies in which they can be isolated and studied.
(3) They show a well defined period-luminosity relation making them
important galactic and extragalactic distance indicators. (4) Mira periods
are related to age and/or metallicity. They are therefore important
traces of galactic structure and kinematics. 

\section{The O-Miras}
Consider first the O-Miras. In the LMC these variables show a narrow 
period-luminosity relation in the infrared. At K (2.2 microns) the
scatter about this relation is only 0.13mag (Feast et al. 1989).
So far as we have been able to tell the slope of this relation is the 
same everywhere (i.e. in the SMC, in globular clusters and for Miras
with individual distances from proper motion companions, Whitelock et al.
1994, Wood 1995, Feast and Whitelock 1999, Feast,
Whitelock \& Menzies 2002). Hipparcos
allowed the zero-point of this relation to be calibrated directly
from Mira parallaxes (Whitelock \& Feast 2000a) and also indirectly
through Miras in globular clusters with 
distances based on Hipparcos subdwarf parallaxes
(Feast et al. 2002).
GAIA will allow one to study this relation in great detail in the
galactic field, provided the necessary ground-based infrared observations
have been made (a programme along these lines has been started at
the South African Astronomical Observatory by 
P.A. Whitelock, F. Marang and the writer). For the Miras with good
coverage of the light curve by GAIA it will also be possible to
discover whether one can replace the infrared photometry
with some of the GAIA photometry. It may be that the GAIA photometry,
together possibly with the spectra, will show evidence for a
period-luminosity-colour relation of which there was a hint in the
infrared photometry (Feast et al. 1989).  

The galactic kinematics of O-Miras are particularly interesting
and Hipparcos led to some some unexpected results.
It has long been known that the galactic kinematics of 
O-Miras are a function of period (e.g. Feast 1963) and this allows the
galactic kinematics of old and intermediate age stars to be
studied as a function of age and/or metallicity in finer detail than is
possible in other ways. Hipparcos improved this discussion in three
ways. (1)The Hipparcos calibration of the PL(K) relation, mentioned
above, allowed one to obtain the distances of many local Miras. (2)
The Hipparcos proper motions could then be used together with (ground-based)
radial velocities to obtain space motions
(Feast \& Whitelock 2000a). Full space motions are
essential if we are to study galactic kinematics free of assumptions. 
(3) The Hipparcos photometry together with ground-based
infrared photometry showed that the short period O-Miras divided
into two groups with different mean colours at the same period
(Whitelock, Marang \& Feast 2000, Whitelock 2002). These two sequences 
were called
the short-period (SP) -red and -blue groups. 
These two groups differ in their kinematic properties and it will
be essential in any future work to distinguish between them. In the 
case of GAIA observations, this may perhaps be done spectroscopically
since the SP-red stars have a later mean spectral type than the SP-blues
at a given period. It may also perhaps be possible using the GAIA
photometry.
On the basis of infrared colours
alone
which show the division into two groups, though less clearly, and on the basis
of kinematics one can show that Miras in globular clusters belong
to the SP-blue group which,
together with longer period Miras, can be called the main Mira sequence. 
These Miras in clusters, and by implication the Miras in the main Mira
sequence generally, lie at the tip of the AGB and are the brightest
objects, bolometrically, of their populations.

What of the SP-red Miras? They have different colours, spectral types
and kinematics from the SP-blue Miras (see Whitelock et al 2000 and Feast
\& Whitelock 2000a for details).Their kinematics associate them with longer
period Miras on the main Mira sequence. At present the best guess is that
they are not yet at the end of their AGB lives but will evolve
into Miras of longer period on the Mira main sequence. There is some 
evidence (Whitelock and Feast 2000a) that at a given period the
SP-red Miras are brighter than the SP-blues. This would be consistent with the
above discussion. Clearly we can anticipate that GAIA will clarify
this whole picture using its combination of parallaxes, proper motions,
radial velocities, spectral types and photometry. This will be a considerable
contribution both to our understanding of AGB evolution and to the study
of Our Own and other galaxies.

When the SP-red stars are omitted, the rest (i.e. the
main Miras sequence,
including the SP-blue variables), show a monotonic dependence of kinematics 
on period.
In particular, the space motions show a smooth change in $V_\theta$ 
(the velocity in the direction of galactic rotation) from  
$133 \pm 19\, \rm km\,s^{-1}$ at a mean period
of 173 days to 
$223 \pm 4\, \rm km\,s^{-1}$ at a period of 453 days (see Feast \& Whitelock 
2000a 
and
Feast 2002). This is for stars which are mostly within about
1 kpc of the Sun. This dependence opens up the possibility of studying
galactic kinematics of homogeneous populations of old and intermediate
age stars on a much finer grid than is otherwise possible. Some
guidance as to the nature of this grid is provided by Miras in globular 
clusters. These are confined to the more metal-rich clusters and to the
shorter period Miras. In addition there is a rather clear period-metallicity
relation for Miras in clusters (see Feast and Whitelock 2000b). To
understand this relation we need to know whether the initial masses
of the Miras in clusters (i.e. the turn-off masses) are a function of
metallicity. This in turn requires us to know whether all the metal-rich
globulars are the same age or whether age depends on metallicity. GAIA
will make a major contribution to settling this issue by providing
much improved estimates of globular cluster distances, either
directly or through subdwarf parallaxes. However it seems worth noting that
this will probably also depend on there being much more detailed ground-based
spectroscopic work to determine chemical abundances in subwarfs both in the
field and in clusters. 

Apart from the globular cluster Miras there is rather little direct
evidence on the initial masses of Miras. R Hya is a member of the Hyades
moving group (Eggen 1985) and this suggests an initial mass of 
$\sim 2 M_{\odot}$.
However this Mira is unusual, its period having varied from about 500 days
to 385 days in the last 340 years
(see e.g. Zijlstra, Bedding \& Mattei 2002), probably due to the star being in
a shell-flashing stage (see below). General arguments based on galactic
kinematics and scale heights (e.g. Olivier, Whitelock \& Marang 2001) 
suggest masses
of $\sim 2 M_{\odot}$ for Miras with periods in the range
$\sim$ 400 to 800 days and $\geq 4 M_{\odot}$ for periods of $\geq 1000$ days.
In the Magellanic Clouds (Nishida et al. 2000) there are intermediate age 
clusters with
turn-off masses of $\sim 1.5 M_{\odot}$ which contain C-Miras
of period $\sim 500$ days. It may
very well be that C- and O-Miras of the same period have the same
initial masses but this has not yet been definitely established. 
We can expect GAIA to make an important contribution to the mass 
question. For instance the space motions from GAIA should allow the
isolation of new moving groups some of which may contain Miras.
We already know of a few Miras with common proper motion companions.
Such objects are potentially important for the determination of limits
to the initial mass of the Mira as well as its luminosity and, 
importantly, its chemical composition which is difficult to derive
directly for a Mira in view of its cool and complex atmosphere.
Incidentally, it is not clear that the Hipparcos catalogue has been
properly searched yet for Miras with proper motion companions.

The dependence of the mean motion of Miras in the direction of
galactic rotation ($V_\theta$) on period shows their importance for
problems of galactic structure and dynamics. A quite unexpected result
was found from the motion of Miras radially outwards from the galactic
centre ($V_{R}$). 
Dividing the Miras into groups according to period, it was found that
for Miras in the groups with mean periods of  
228 days or more there is
a small net outward motion of a few $\rm km\,s^{-1}$. The significance of 
this is
not at present very clear. However a rather startling result was
found for the shortest period group (mean period 173 days) when the
SP-red varables were omitted.
These stars show a marked asymmetric drift 
($V_{\theta} = 133\pm 19 \, \rm km\,s^{-1}$, compared
with a circular velocity of $231 \, \rm km\,s^{-1}$
(Feast \& Whitelock 1997)).
Whilst this group is small (18 stars)
it shows rather clearly a mean outward motion, $V_{R}$, of
$75 \pm 18 \, \rm km\,s^{-1}$. In addition
the individual space motions show that there is a rather good correlation
between $V_{\theta}$ and $V_{R}$ in this group and all 11 stars
which have an asymmetric drift greater than 
$65 \, \rm km\,s^{-1}$ having positive
values of $V_{R}$ (see 
fig. 1 and Feast and Whitelock 2000a, Feast 2002).

\begin{figure}
\plotone{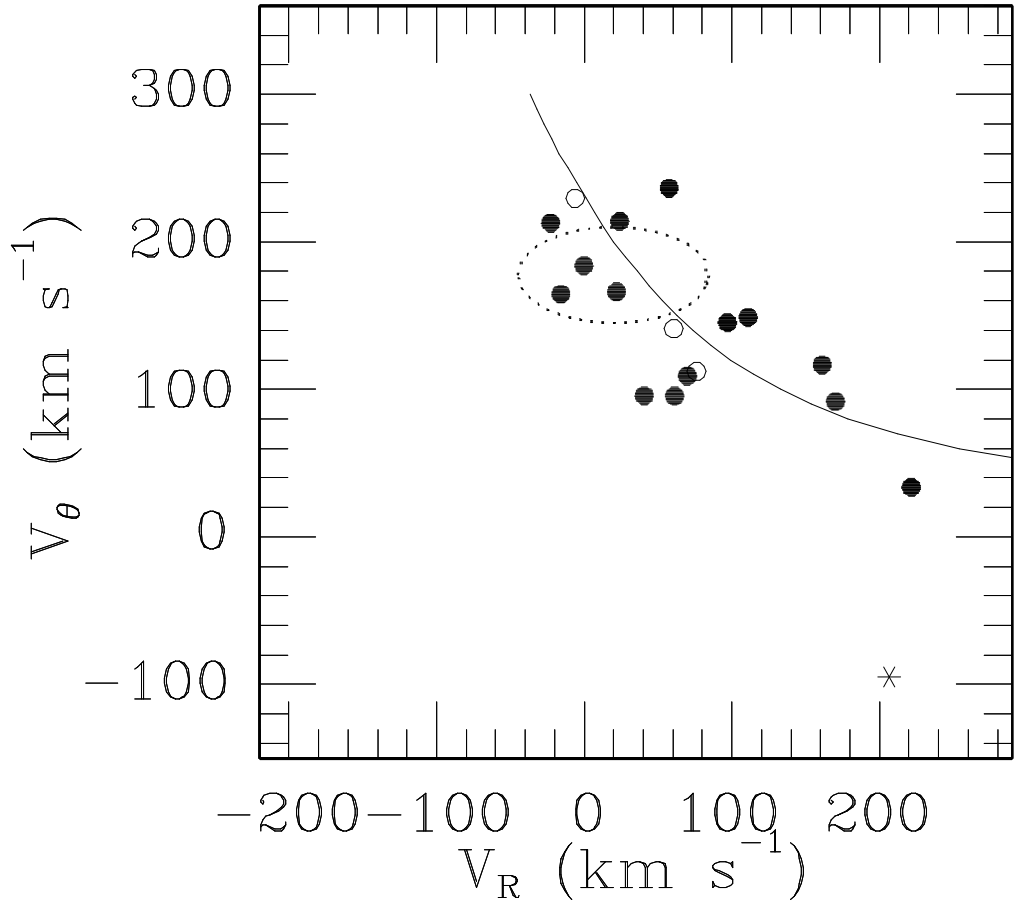}
\caption{ The correlation between $V_{\theta}$ and $V_{R}$ for short period,
blue-sequence, Miras. The solid line is from the simple model of
Feast \& Whitelock 2000a. The dotted oval shows the region occupied
by the ``Hercules"stream (see Fux 2001 and Feast 2002).
The asterisk is for S Car which is on a highly eccentric retrograde orbit.}
\end{figure}

Since many of these stars, all of which are within about 1 kpc of the Sun,
are on highly eccentric orbits and will pass through, or close to, the
Galactic Bulge, it seems likely that they are members of a bar-like
distribution. 
It is known (Whitelock \& Catchpole 1992) that the Miras in the
Bulge itself belong to a central bar structure.
A simple model would fit this suggestion for the local Miras although the
possibility that they are part of a stream due to an infalling 
satellite galaxy, cannot be ruled out (see Feast and Whitelock 2000a and
Feast 2002 for details). It is evident that a detailed study of this
phenomenon requires space motions for short period Miras over a wide region
of the Galaxy. This will be possible with GAIA parallaxes, proper motions,
radial velocities and periods. It remains to be seen whether auxilliary
data will be require to separate out the SP-red Miras which have 
different kinematics.

Related to this problem is that of
the Galactic Bulge itself where, 
as just mentioned, Miras show a bar-like distribution.
The Bulge contains Miras of a wide range of period and GAIA should
allow the kinematics 
and structure of the Bulge to be explored 
in detail
as a function of period using
positions and space motions.

\section{The C-Miras and S-type Miras}
An extensive discussion of the carbon-rich Miras is beyond the
scope of this talk. However the parallaxes, proper motions, radial
velocities, periods and spectra which GAIA will provide
should answer a number of outstanding problems connected with these
objects. The space motions of C-Miras are of importance in their
own right and will show how closely these stars are related to
the O-Miras of the same (or a different) period. Whilst it seems likely 
that C-Mira evolve from O-Miras this is not yet certainly established.
If indeed this is their evolutionary path, it is still uncertain
whether they change period in the process.
  
Space motions derived by GAIA will also be important for the understanding
of the Miras with spectra of type S. These show strong zirconium oxide
bands and lines of s-process elements are enhanced. These stars, which are
oxygen-rich, are generally treated together with the
O-Miras which have strong TiO bands. There is no clear dividing
line between Miras with M type  and those with S type spectra, the
classes merging into one another
(the MS stars). The extensive spectral and other
data on O-Miras from GAIA may make it possible to divide them into
groups which are more clearly defined and should lead to a much
clearer understanding of this whole group of stars, their place
in stellar evolution and their use in Galactic and extragalactic
astronomy.

\section{The Low Amplitude, Semiregular Variables}
  In this review a detailed discussion of the low-amplitude semiregular cool
variables (the SR stars) has been omitted. However GAIA will almost 
certainly
bring considerable order into this field which at present is
in a rather unsatisfactory and confused state. In globular clusters,
SR variables evolve with increasing period and luminosity 
(see e.g. Whitelock
1986, Feast 1989). They may change pulsation mode as they do so.
In the LMC there are several sequences of semiregular,
or low amplitude,  variables in
the period - luminosity plane, probably indicating several different
pulsation modes (Wood 2000). The GAIA data will presumably reveal
similar sequences in our Own Galaxy (though infrared photometry of
the relevant objects will probably be necessary
to do this). With this done the
kinematics, based on space motions, of the semiregulars divided
according to particular period-luminosity sequence and to period, will
help establish their evolutionary relationship. In view of the somewhat
surprising results discussed above for the short period O-Miras, it
seems very desirable that any kinematic analysis of these stars uses
the full space and velocity co-ordinates. The spectra together with the
photometry also opens up a largely unexplored area.

\section{Deviations from  the Mira Period-Luminosity Relation}
If our ideas based on the Magellanic Clouds, globular clusters and
the Hipparcos results on local Miras, are correct, it
can be anticipated that in any volume of space 
in the Galaxy most of the regular,
large amplitude, cool variables measured by GAIA will be Miras on the
PL relation (either O- or C- Miras). 
Together with infrared photometry GAIA will establish the slope of
this relation and its zero-point.
The parallaxes will also 
enable one to find Mira or Mira-like stars that lie off the PL relation. 
Some stars of this kind are expected and their 
luminosities, space motions and spectra will be of great interest. 

The SP-red Miras which seem to lie above the main PL relation were mentioned
in section 2. 
But there are other types of Mira or Mira-like stars as well which are 
expected to lie off the PL relation.
Towards the end of their AGB evolution, stars with initial masses in the
range 4 to 6 solar masses can undergo Hot-Bottom-Burning (HBB). In this
process the base of the H-rich convective envelope dips into the H-burning 
shell.
The luminosity can then rise above that predicted by the classical
relation between core-mass and luminosity. Carbon is burned to nitrogen
and the beryllium transport mechanism results in an overabundance of
lithium at the surface. Whitelock (2002, see also Whitelock \& Feast 2000b
and references there) has pointed out that most of the AGB variables
in the Magellanic Clouds in which Smith et al. (1995) found lithium to be
strong lie above the Miras PL relation. 
The frequency of occurrence of
these stars and their kinematics are obviously of great interest since they
gives clues to lifetimes and initial masses. Whitelock (2002) 
points out that the HBB
phenomenon can explain a curious result found for LMC Miras. The
evidence, which she summarizes, suggests that most LMC Miras
with relatively thick dust shells and periods in the range
420 to 1300 days lie on 
an extension of the (bolometric) PL relation defined by shorter period
Miras.
These longer-period stars
may be of sufficiently low mass that they never undergo HBB. However
she points out that the few Miras with periods greater than 420 days
and rather thin dust shells which seemed to indicate a break
in the PL relation at that period 
in earlier work
(Feast et al. 1989) may in fact be higher mass stars in a
HBB phase. A full understanding of these phenomena is important not only 
for stellar evolution theory but also for the use of Miras as
extragalactic distance indicators. For instance Whitelock (2002)
suggests that the 641 day variable in IC1613 (Kurtev et al 2001) which lies
well above the Mira PL may be in the HBB phase.

It has long been thought that the slow changes in period of
some Miras (e.g. R Hya, see above) are due to the star undergoing
helium shell flashing (thermal pulsing) (Wood \& Zarro 1981). This is expected 
to be 
accompanied by
changes in luminosity (e.g. Iben \& Renzini 1983). Thus we may expect
to find variables both above and below the PL relation due to this
phenomenon. However the behaviour of R Hya is complex and the thermal pulse
model has recently been challenged by Zijlstra et al. (2002) who suggest 
instead
an envelope relaxation model.

Finally one should mention the possibility of finding dust-enshrouded
OH/IR Miras below the PL relation. There is some evidence for such stars in 
the region of the Galactic Centre (e.g. Blommaert et al. 1998, Wood et al.
1998) but the bolometric luminosities of these stars are difficult
to estimate accurately.
\section{Red Supergiant Variables}
  Especially near the Galactic plane, GAIA will measure red supergiant
variables. These objects show 
bolometric or infrared period-luminosity relations in, for instance,
the Magellanic Clouds and M33 (Kinman, Mould \& Wood 1987, 
Mould et al. 1990,
Feast 1992). More recently, and of more direct relevance for GAIA, 
period- luminosity relations in the I-band have been established
in Per OB1, LMC, M33 and M101 (Pierce, Jurcevic \& Crabtree 2000, 
Jurcevic, Pierce \& Jacoby 2000).
In the I-band these red supergiant variables are about three magnitudes
brighter than the classical Miras at a given period. The range in periods
in Per OB1 suggests that the PL relation there is an evolutionary sequence
in contrast to the Mira PL relation which is a mass/metallicity sequence,
as discussed above. The PL relation established by Pierce et al. has a 
considerable
RMS scatter (0.42 mag). This may be partly observational.
It remains to see whether this PL relation is influenced
by inital mass,age, or metallicity .
These stars should be detected by GAIA to large distances even in the
galactic plane and should be excellent tracers of the distribution
and kinematics of young objects. The radial velocity
component will be vital to study the kinematics properly. 
\section{Some Special Challenges}
There will be a number of special challenges to be met in the interpretation
of GAIA observations of Mira variables. One challenge is of course that 
we are moving into a little explored area. For instance, there are few,
if any Miras that have been studied systematically round their light cycles
for spectroscopic and radial velocity variations in the spectral
region of interest to GAIA. The complex structure of Mira atmospheres 
and its variation with phase is not yet properly understood. References
to high resolutions studies
and their interpretation, from the early work of Merrill to the present
are conveniently summarized by Alvarez et al. (2001); see also the
summary of Lebzelter and Hinkle (2002). In the optical and infrared
regions doubling of absorption lines is seen. Typically the lines
are split by $\sim 20 \, \rm km\, s^{-1}$
in the optical region. At resolutions too low to show the 
splitting clearly, the variation in 
measured absorption line velocity round
the cycle is probably $\sim 10 \, \rm km\, s^{-1}$ or less, though there 
have been
no studies 
in the optical region
as extensive those of Joy (1954) on Mira Ceti itself. There are
variations of absorption line velocities with excitation potential,
presumably due to the different depths of formation of the lines. The line 
doubling is thought to be due to shock waves in the atmosphere. These
are also believed to excite the emission lines seen at some phases. 
In the past there has been some uncertainty
about the definition of the actual radial velocity of a Mira in space.
This is now generally taken to be the mean of the velocities of the two
OH maser peaks which are found in some Miras. From a large body of data,
generally involving only a few optical velocities for any given Mira, it is 
found that the mean absorption line velocity
is too positive by $4 \, \rm km\, s^{-1}$ 
with some dependence on period (Feast and Whitelock 2000a). This 
offset may be due to
most of the optical velocities having been obtained in the brighter half
of the light cycle.
Perhaps
of more relevance, so far as GAIA spectroscopy is concerned, 
is the fact that some Miras, at least, show emission in
the CaII infrared lines near maximum light.  
At least in some cases
these lines have
inverse P-Cygni profiles with velocity separations of $20-30 \, \rm km\,s^{-1}$
(Merrill 1934, 1960).
Clearly, this  must be taken into account in deriving mean velocities
and space motions
using these lines. The systematic study of the CaII lines round the
light cycles of many Miras will be very revealing of the complex
dynamics of Mira atmospheres.

A challenge, which affects particularly the astrometry, is the large 
angular size of the Miras.
The diameters of Miras are a strong function of wavelength 
(e.g. Labeyrie et al. 1977). In a broad optical band, which is probably
relevant to GAIA, the Mira R Leo and 
the Mira-like (SRa) star W Hya have mean angular diameters of
74 and 84 mas. These stars do not have
circular symmetry
and show evidence for asymmetric light distributions over the discs,
possibly due to large star-spots
(Lattanzi et al. 1997 and references there). 
The diameters of these stars are greater than the expected
point spead function of GAIA. 
They are also nearly
a factor of ten greater than their Hipparcos parallaxes 
(or the parallaxes derived from the PL(K) relation of Feast et al. (2002)). 
Thus the diameters of Miras  are always larger than their parallaxes and 
astrometry may
be affected by motions of the photocentre of the star due to changing
shape or light distribution. Such effects
might be incorrectly interpreted as due to motion in a binary.
Whilst this is a serious concern, 
it should be noted that one of the fastest growing areas in astronomy
at the present time is stellar interferometry. One may hope that by
the time GAIA results are available, interferometry will have led to a
good understanding of the shapes and surface structure of Miras and, in
particular, the time scale on which these change, something on which
we have little information at present.
\section{Conclusions}
The combination of GAIA astrometry, radial velocities, spectra, 
photometry and periods will have
a profound and unique effect on our understanding of the nature and evolution
of cool variable stars, particularly those on the AGB. Also the kinematics
of these objects will allow us to study the structure and evolution of Our
Galaxy in a way previously impossible. Hipparcos already led to surprises
in this area and gave some foretaste of what GAIA will accomplish.

It would be rather valuable if ground-based observers gave some thought now
to parallel programmes which would be completed by the time the
results from GAIA become available. In the optical/infrared field some
obvious programmes in this category are; interferometry, intensive
infrared photometry and high resolution optical spectroscopy. 
\acknowledgments
  I would like to thank Patricia Whitelock for discussions and advice.

\end{document}